\documentclass[3p,times,twocolumn,sort&compress]{elsarticle}

\usepackage{epstopdf}
\usepackage{amsmath}
\usepackage{amssymb}
\usepackage{amsfonts}
\usepackage{graphicx}
\usepackage{subfig}
\usepackage{natbib}
\usepackage{color}
\usepackage{lineno}
\journal{Physics Letters B}

\begin{document}

\begin{frontmatter}

\title{First observation of two hyperfine transitions in antiprotonic $^3$He}

\author[1]{S.Friedreich}
\ead{susanne.friedreich@oeaw.ac.at}
\author[2,3]{D.~Barna}
\author[4]{F.~Caspers}
\author[2]{A.~Dax}
\author[2]{R.S.~Hayano}
\author[2,5]{M.~Hori}
\author[3,6]{D.~Horv\'ath}
\author[1]{B.~Juh\'asz}
\author[2]{T.~Kobayashi}
\author[1]{O.~Massiczek}
\author[5]{A.~S\'ot\'er}
\author[2]{K.~Todoroki}
\author[1]{E.~Widmann}
\author[1]{J.~Zmeskal}

\address[1]{Stefan Meyer Institute for Subatomic Physics, Austrian Academy of Sciences, Boltzmanngasse 3, A-1090 Vienna, Austria}
\address[2]{Department of Physics, University of Tokyo, 7-3-1 Hongo, Bunkyo-ku, Tokyo 113-0033, Japan}
\address[3]{KFKI Research Institute for Particle and Nuclear Physics, H-1525 Budapest, PO Box 49, Hungary}
\address[4]{CERN, CH-1211 Geneva, Switzerland}
\address[5]{Max-Planck-Institut f\"{u}r Quantenoptik, Hans-Kopfermann-Strasse 1, D-85748 Garching, Germany}
\address[6]{Institute of Nuclear Research of the Hungarian Academy of Sciences, H-4001 Debrecen, PO Box 51, Hungary}

\begin{abstract}
We report on the first experimental results for microwave spectroscopy of the hyperfine structure of $\overline{\text{p}}^3$He$^+$. Due to the helium nuclear spin, $\overline{\text{p}}^3$He$^+$ has a more complex hyperfine structure than $\overline{\text{p}}^4$He$^+$, which has already been studied before. Thus a comparison between theoretical calculations and the experimental results will provide a more stringent test of the three-body quantum electrodynamics (QED) theory.  
Two out of four super-super-hyperfine (SSHF) transition lines of the $(n,L)=(36,34)$ state were observed. The measured frequencies of the individual transitions are $11.12559(14)$~GHz and $11.15839(18)$~GHz, less than 1~MHz higher than the current theoretical values, but still within their estimated errors. Although the experimental uncertainty for the difference of these frequencies is still very large as compared to that of theory, its measured value agrees with theoretical calculations. This difference is crucial to be determined because it is proportional to the magnetic moment of the antiproton.
\end{abstract}


\end{frontmatter}

\section{Introduction}
\begin{linenumbers}
It was observed for the first time at KEK in Japan in 1991~\cite{Iwasaki:91} that antiprotons stopped in helium can survive for several microseconds.

If an antiproton is approaching a helium atom at its ionization energy (24.6~eV) or below, the antiproton can eject one of the two electrons from the ground state of the helium atom, replace it and thus get captured. This exotic, metastable three-body \textit{antiprotonic helium}, i.e. $\overline{\text{p}}$He$^{+}$, consists of one electron in the ground state, the helium nucleus and the antiproton~\cite{Yamazaki:93,Yamazaki:02,Hayano:2007}. The atoms occupy circular states with $L$ close to $n$, where $L$ is the angular momentum quantum number and $n$ the principal quantum number. The electron remains in the ground state. The antiproton is, due to its high mass, most likely to be captured into states with high angular momentum, i.e. $n = n_0 \equiv \sqrt{M^{*}/m_{\text{e}}} \sim 38$, $M^{*}$ being the reduced mass of the system.

About 97\% of these exotic atoms find themselves in states dominated by Auger decay and ionize within a few nanoseconds because of the Auger excitation of the electron. Afterwards, the remaining antiprotonic helium ion undergoes Stark mixing due to the electric field of the surrounding helium atoms. The antiprotons then annihilate within picoseconds with one of the nucleons of the helium nucleus because of the overlap of their wave functions. 

Only 3\% of the antiprotonic helium atoms remain in metastable, radiative decay-dominated states. In this case, the change of orbital angular momentum in the Auger transition is large and thus Auger decay is suppressed. Consequently, these states are relatively long lived, having a lifetime of about 1-2~$\mu$s. This time window can be used to do microwave spectroscopy measurements.
\end{linenumbers}

\section{Hyperfine structure of antiprotonic helium}
\begin{linenumbers}
The interaction of the magnetic moments of its constituting particles gives rise to a splitting of the $\overline{\text{p}}^{3}$He$^{+}$ energy levels. The coupling of the electron spin $\vec{S}_{\text{e}}$ and the orbital angular momentum of the antiproton $L$ leads to the primary splitting of the state into a doublet structure, referred to as \textit{hyperfine (HF) splitting}. The quantum number $\vec{F}=\vec{L}+\vec{S}_\text{e}$ defines the two substates as $F_+=L+\frac{1}{2}$ and $F_-=L-\frac{1}{2}$. The non-zero spin of the $^{3}$He nucleus causes a further, so-called \textit{super-hyperfine (SHF) splitting}, which can be characterized by the quantum number $\vec{G}=\vec{F}+\vec{S}_{\text{h}}=\vec{L}+\vec{S}_{\text{e}}+\vec{S}_\text{h}$, where $\vec{S}_\text{h}$ is the spin of the helium nucleus. This results in four SHF substates. At last, the spin-orbit interaction of the antiproton orbital angular momentum and antiproton spin $\vec{S}_{\bar{\text{p}}}$ in combination with the contact spin-spin and the tensor spin-spin interactions between the particles result in a further splitting of the SHF states into eight substates -- as illustrated in Fig.~\ref{fig:He3_lMWl} --, which we call \textit{super-super-hyperfine (SSHF) splitting}. This octuplet structure can be described by the quantum number $\vec{J}=\vec{G}+\vec{S}_{\bar{\text{p}}}=\vec{L}+\vec{S}_{\text{e}}+\vec{S}_{\text{h}}+\vec{S}_{\bar{\text{p}}}$. Even though the magnetic moment of the antiproton is larger than that of the $^{3}$He nucleus, the former has a smaller overlap with the electron cloud. Therefore, it creates a smaller splitting.
\linebreak

In $\overline{\text{p}}^4$He$^+$, however, where the $^{4}$He nucleus has zero spin, only a quadruplet structure is present. The hyperfine structure of the $(n,L)=(37,35)$ state of $\overline{\text{p}}^4$He$^+$ was already extensively studied. Through comparison of the experimental results to state-of-the-art three-body QED calculations a new experimental value for the spin magnetic moment of the antiproton was obtained as $\mu_{\text{s}}^{\bar{\text{p}}}=2.7862(83)\mu_{\text{N}}$~\cite{Pask:09}, where $\mu_{\text{N}}$ is the nuclear magneton. This is more precise than the previous measurement by Kreissl et al.~\cite{Kreissl:88}. The agreement with $\mu_{\text{s}}^{\text{p}}$ was within 0.24\%~\cite{Widmann:02, Pask:08}.
\linebreak

New microwave spectroscopy measurements with $\overline{\text{p}}^3$He$^+$ were started, studying the state $(n,L)=(36,34)$. It was the first attempt to measure the microwave transition frequencies of antiprotonic $^3$He. Transitions between the SSHF states can be induced by a magnetic field oscillating in the microwave frequency range. Due to technical limitations of the microwave input power, only the transitions which flip the spin of the electron can be measured. There are four such "allowed" SSHF transitions for the $(n,L)=(36,34)$ state of $\overline{\text{p}}^3$He$^+$, of which we report on two:

\begin{align}     
\nu_{\text{HF}}^{--}~:~J^{---}=L-\frac{3}{2}~\longrightarrow ~J^{+--}=L-\frac{1}{2}   \nonumber   \\
\nu_{\text{HF}}^{-+}~:~J^{--+}=L-\frac{1}{2}~\longrightarrow ~J^{+-+}=L+\frac{1}{2}     
\label{MWtrans_He3}         
\end{align}

The interest in $\overline{\text{p}}^3$He$^+$ arose from its more complex structure due to the additional coupling of the nuclear spin with the antiproton orbital momentum. Such a measurement will allow a more rigorous test of theory. 
The theoretical calculations have been developed by two different groups~\cite{Bakalov:98,Korobov:01,Kino:03APAC}. The hyperfine structure for $\overline{\text{p}}^3$He$^+$ has been calculated by V.~Korobov~\cite{Korobov:06} with the most accurate variational wave functions.
\end{linenumbers}

\section{Laser-microwave-laser spectroscopy}
\label{sec:method}
\begin{figure}
\centering
\includegraphics[scale=.28, trim=0 0 0 0,clip]{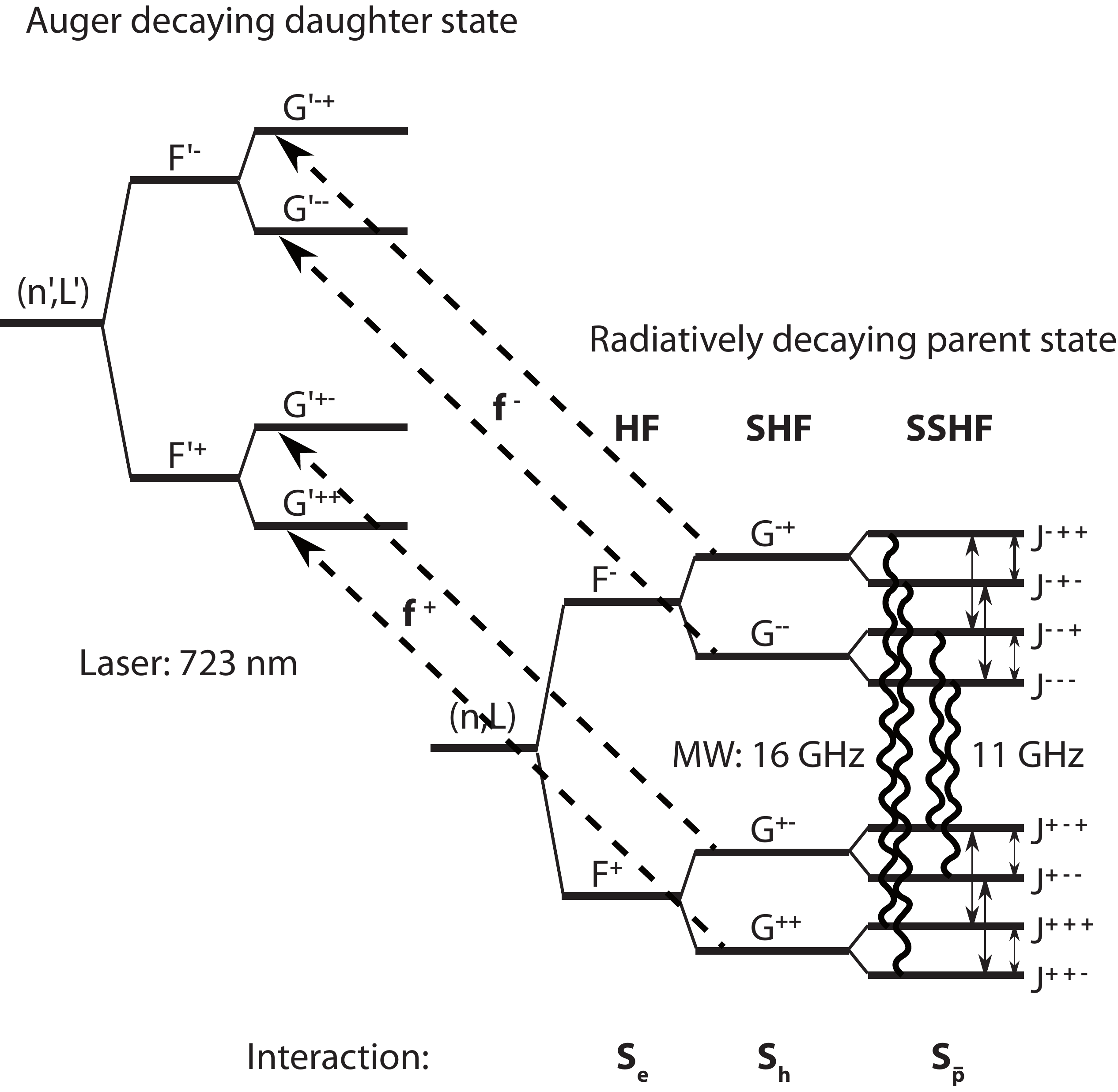} 
\caption{\small{A schematic drawing of the laser-microwave-laser method. The dashed arrows indicate the laser transitions between the SHF levels of the radiative decay-dominated state $(n,L)=(36,34)$ and the Auger decay-dominated state $(n,L)=(37,33)$ of $\bar{\text{p}}^{3}$He$^{+}$. The wavy lines illustrate the microwave-induced transitions between the SSHF levels of the long-lived state.}} \label{fig:He3_lMWl}
\end{figure}

\begin{linenumbers}
The experimental technique is a three-step process, referred to as \textit{laser-microwave-laser spectroscopy} (Fig.~\ref{fig:He3_lMWl}). 

After antiprotonic helium is formed, the atoms in the hyperfine substates are all equally populated. Therefore at first a population asymmetry between the SSHF states of the measured radiative decay state $(n,L)$ needs to be created. This depopulation is induced by a short laser pulse, which transfers the majority of the antiprotons from one of the HF states of the radiative decay-dominated, metastable parent state to the Auger decay-dominated, short-lived daughter state ($f^+$ transition in Fig.~\ref{fig:He3_lMWl}). The bandwidth of the laser is narrow enough so that the $f^-$ transition is not excited. Thus the antiprotons in the other HF state are not affected, which results in the desired population asymmetry. The antiprotons in the short-lived daughter state annihilate within a few nanoseconds. Afterwards, a microwave frequency pulse, tuned around the transition frequency between two SSHF ($\overline{\text{p}}^{3}He^{+}$) substates of the parent state, is applied to the antiprotonic helium atoms. If the microwave field is on resonance with one of the SSHF transitions, this will cause a population transfer and thus a partial refilling of one of the previously depopulated states. Then, a second laser pulse is applied again to the same transition ($f^+$) as before, which will again result in subsequent Auger decay of the transferred atoms and annihilation of the antiprotons. Thus the number of annihilations after the second laser pulse will be larger if more antiprotons were transferred by the microwave pulse. 
\linebreak

The annihilation decay products -- primarily charged pions, but also electrons and positrons -- resulting from the decay of the daughter state after the two laser pulses, are detected by two Cherenkov counters (see Section~\ref{sec:expset}).

\begin{figure}
\centering
\includegraphics[scale=.42, trim=0 0 0 0,clip]{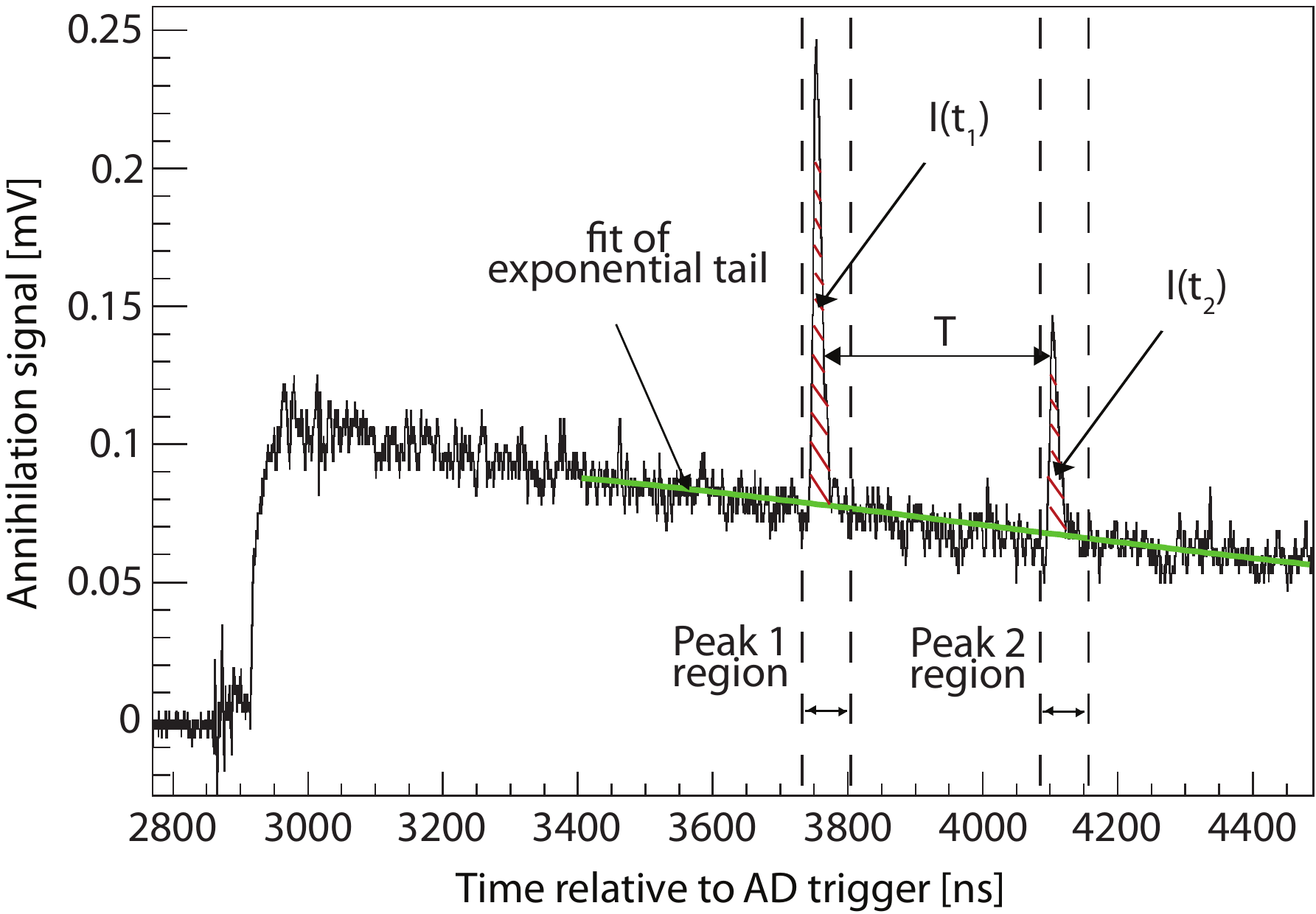} 
\caption{\small{Part of the analog delayed annihilation time spectrum (ADATS) with the two laser-stimulated annihilation peaks against the exponentially decaying background of the metastable cascade. $T$ denotes the delay time between the two laser pulses. The photomultipliers are gated off during the initial $\overline{\text{p}}$ pulse arrival~\cite{Hori:2003}. Thus the prompt peak is cut off and only the annihilations due to the metastable state depopulation are recorded.}} \label{fig:Adats}
\end{figure}

Prior to the first laser-induced population transfer a large annihilation peak (prompt) is caused by the majority of $\overline{\text{p}}\text{He}^{+}$ atoms which find themselves in Auger decay-dominated states and annihilate within picoseconds after formation. At later times, this peak exhibits an exponential tail due to $\overline{\text{p}}\text{He}^{+}$ atoms in the metastable states cascading more slowly towards the nucleus. This constitutes the background for the laser-induced annihilation signals. 
As mentioned above, the daughter state has a very short lifetime of $\sim$10~ns and thus the population transfer is indicated by a sharp annihilation peak against the background during the two laser pulses. The area under these peaks is proportional to the population transferred to the Auger decay-dominated state. This spectrum is called an \textit{analogue delayed annihilation time spectrum} or ADATS. The spectrum with the two laser-induced peaks super-imposed on the exponential tail is displayed in Fig.~\ref{fig:Adats}. 
\linebreak

Since the intensity of the antiproton pulse fluctuates from shot to shot, the peaks must be normalised by the total intensity of the pulse (total). This ratio is referred to as \textit{peak-to-total}. The peak-to-total (ptt) corresponds to the ratio of the peak area ($I(t_1)$ or $I(t_2)$) to the total area under the full spectrum (see Fig.~\ref{fig:Adats}. If the second laser annihilation peak is further normalised to the first one, the total cancels out. The frequencies of the two SSHF transitions can now be obtained as distinct lines by plotting $\frac{I(t_2)}{I(t_1)}$ as a function of the microwave frequency. The ratio $\frac{I(t_2)}{I(t_1)}$ is largely independent of the intensity and position of the antiproton beam. The height of the microwave spectrum lines depends on the time delay between the two laser pulses and collisional relaxation effects which are estimated to be 1~MHz at 6~K~\cite{Korenman:2010}.


For these measurements the two pulsed lasers were fixed to a wavelength of 723.877~nm (see Section~\ref{sec:res}), with a pulse length of 10-12~ns, to induce the $f^{+}$ laser transition between the $(n,L)=(36,34)$ and the $(n',L')=(37,33)$ state. The laser fluence was in the range of 20-40~mJ$/$cm$^{2}$, the laser waist $\sim$5~mm, leading to a depletion efficiency of about 50\%.
There are several limitations to the choice of the measured state, such as availability of a laser source in the required frequency range or the splitting of the transitions between the HF states of the daughter and the parent state. The laser transition between the $(n,L)=(36,34)$ and the $(n',L')=(37,33)$ state was chosen because it is easily stimulated and the primary population is large, thus leading to a large signal. The captured fraction of antiprotons for the measured metastable state $(n,L)=(36,34)$ is (3-4)$\times 10^{-3}$~\cite{Hori:02}.
\end{linenumbers}

\section{Experimental setup}
\label{sec:expset}
\begin{figure}
\centering
\includegraphics[scale=0.38, trim=50 0 0 0,clip]{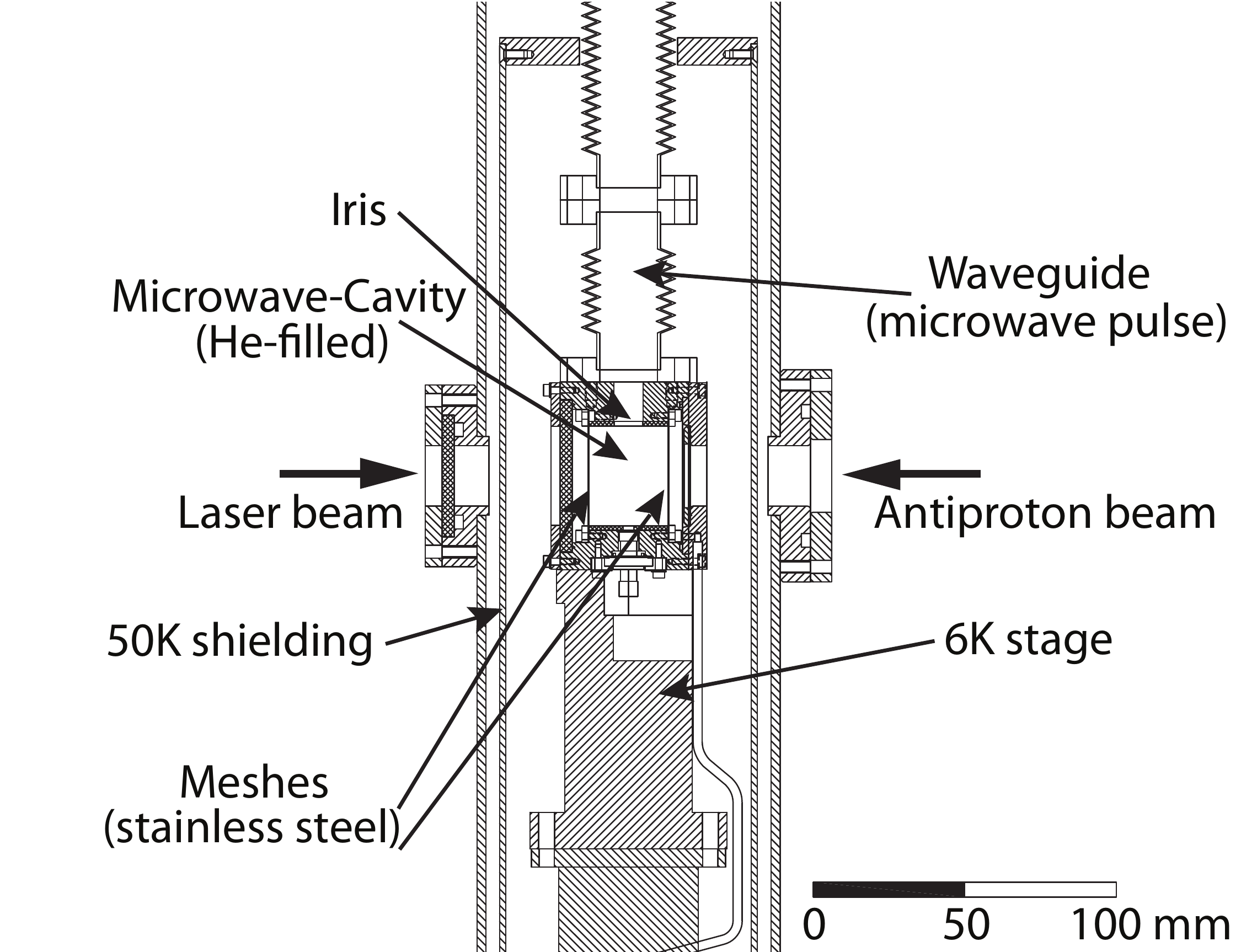}
\caption{\small{Drawing of the central part of the experimental setup, a cross-section of the cryostat.}}
\label{fig:NewCryostat}
\end{figure}

\begin{linenumbers}
The antiprotons for the experiment were provided by the AD (Antiproton Decelerator) at CERN~\cite{Maury:1997}, with a pulsed beam of (1-3)$\times 10^7$ antiprotons at an energy of 5.3~MeV and a repetition rate of about 100~s. The particles were stopped in a helium gas target, cooled down to a temperature of about 6~K. The usual gas pressure was in the range of 150-500~mbar. The gas target was a cylindrical chamber, which also acted as a microwave resonance cavity. In order to measure the annihilation decay products two Cherenkov counters were mounted around the target volume. The resulting photons are detected by photomultipliers, which are gated off during the initial $\overline{\text{p}}$ pulse arrival~\cite{Hori:2003}. The microwave pulse was synthesized by a vector network analyzer (VNA, Rhode \& Schwarz ZVB20) and amplified by a traveling wave tube amplifier (TWTA, TMD PTC6358). A waveguide system then transmitted the microwave pulse of $\sim$20~$\mu$s to the cavity. Fig.~\ref{fig:NewCryostat} gives an overview of the central part of the setup.
\linebreak

The experimental method and the general design of the setup were the same as for $\overline{\text{p}}^4$He$^+$~\cite{Sakaguchi:04}. However, a new cryostat with compressor-based cooling system was built, which led to improvements of the operation and more efficient use of the measurement time. The microwave cavity was now cooled directly by mounting it on a coldhead. Only the cavity was filled with the helium gas and by means of the coldhead cooled down to about 6~K. Liquid nitrogen and liquid helium were no longer needed for the cooling process. The temperature stabilization was much faster compared to the old system. The cryostat could be operated continuously and thus saved $\sim$10\% of beamtime previously needed for refilling of the cryogenic coolants. 
\linebreak

Out of the four "allowed" SSHF transitions of the state $(n,L)=(36,34)$ of $\overline{\text{p}}^3$He$^+$, two of them lie within 32~MHz from each other. Therefore, these two transitions can be measured with a single cavity with a resonance frequency of 11.14~GHz, which is in the middle between the two SSHF transition frequencies. For the other two transitions another cavity with a resonance frequency of 16.13~GHz will be used in the future.
\linebreak

The cavity was built of brass, which has proven to be the optimum combination of machinability and thermal conductivity and is further non-magnetic. The central frequency of a cylindrical microwave cavity is defined by its dimensions -- length $l$ and radius $r$. The cavity used for the measurements of the 11~GHz transitions has a radius of 16.19~mm and a length of 26.16~mm. It is oriented in a way that the antiproton beam and the laser beam enter along the axis of the cavity. Two stainless steel meshes (transmissibility $>$90\%, 250~$\mu$m thick, wire thickness 0.05~mm, wire distance 0.75~mm) on both faces of the cylinder confine the radio-frequency field inside the cavity, still allowing the laser and the antiprotons to enter the target.
\linebreak

Radius and length -- and their ratio $(\frac{2r}{l})^2$ -- also determine which field modes can resonate inside the cavity and at which frequency. For this experiment the transverse magnetic mode $\text{TM}_{110}$ for a cylindrical cavity has been chosen, which is parallel to the axis of the cylindrical cavity and independent of the cavity length, i.e. on the direction of the antiproton beam. It is useful to have a homogeneous field over a large range in the direction of the antiproton beam in order to be less sensitive to the stopping region of the antiprotons in the gas. It is also desirable that the field distribution is uniform over the region where the laser is applied -- which is usually smaller than the diameter of the slowed down antiproton beam to achieve sufficient laser power density. Further, the size of the homogeneous resonating field in transverse direction to the beam should be comparable to the stopping distribution of the antiprotons. 
\linebreak

A broad resonance of the cavity is necessary to allow scanning over a large enough frequency range in order to measure both SSHF transitions with the same cavity at an equal power level~\cite{Sakaguchi:04}. The field is measured with a small ($\sim$2~mm) pin antenna opposite to the waveguide input. Thus the power in the cavity was monitored and the input power adjusted for every frequency point, with a power fluctuation of $\sim$16\% over the frequency range. The only drawback is that considerably high input power, i.e. up to 200~W, might be required. 

To obtain a sufficiently broad resonance with a FWHM of at least 100~MHz the cavity is over-coupled to the waveguide system through an iris, a rectangular aperture in the cavity. The iris of the 11~GHz target has a size of 7$\times$8~mm, the longer side in radial direction. This way the resonance width of the cavity was $\sim$140~MHz. These are fixed parameters of the setup. When designing the target, the width of the resonance can be optimized by changing the iris dimensions -- and consequently also the central frequency shifts, which can be readjusted by changing the radius. The length may also have to be adapted in order to exclude interferences with other field modes. In particular the iris size is crucial for a successful measurement since it may cause polarization degeneration of the mode in the cavity.   
The whole microwave part of the setup was designed using the High Frequency Structure Simulator (HFSS) Software~\cite{HFSS}.

\end{linenumbers}

\section{Results}
\label{sec:res}
\begin{figure}
\centering
\includegraphics[scale=.35, trim=300 150 350 200,clip]{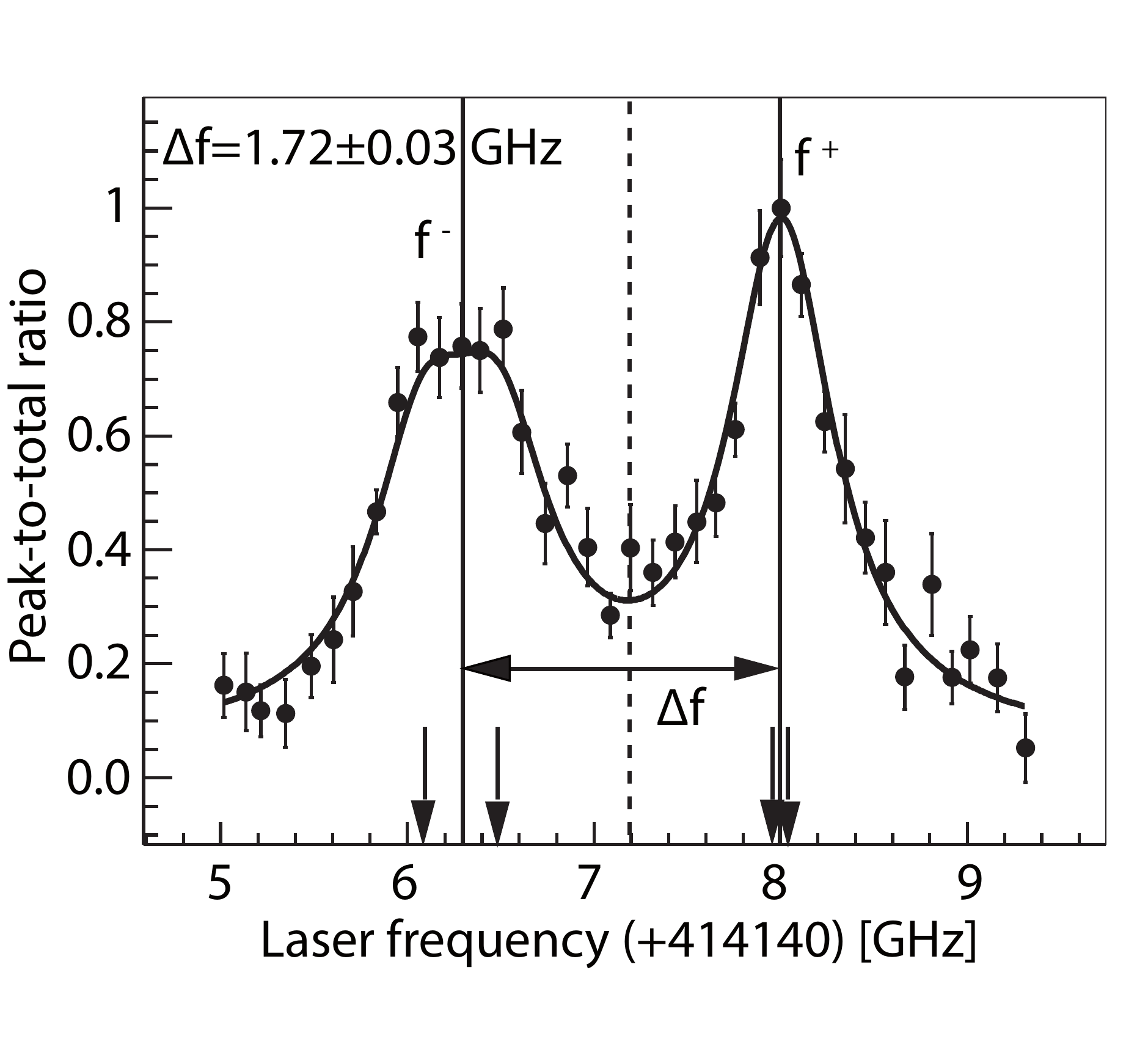} 
\caption{\small{Laser resonance profile for the $(n,L)=(36,34)$ state of $\overline{\text{p}}^{3}$He$^{+}$, displaying the two laser transitions $f^{+}$ and $f^{-}$ between the HF states of the parent and the daughter state, at a target pressure of 250~mbar. The peaks are fitted with four Voigt functions referring to the four "allowed" E1 transitions between the SHF states of the parent state (refer to Fig.~\ref{fig:He3_lMWl}). The arrows indicate the corresponding theoretical transition frequencies.}}
\label{fig:Laserscan}  
\end{figure}

\begin{linenumbers}
First, a scan over the laser frequency range was done to determine the frequency offset and the splitting of the two HF lines (see Fig.~\ref{fig:Laserscan}) to ensure that only one of the two hyperfine levels of the $(n,L)=(36,34)$ state is depopulated by laser stimulation. 

The splitting is $\Delta f = 1.72 \pm 0.03$~GHz, similar to the transition at $\sim$726.090~nm in $\overline{\text{p}}^4$He$^+$, with a splitting of $\Delta f = 1.75 \pm 0.01$~GHz. Due to the different SSHF energy level spacings, one of the laser transition peaks has a lower amplitude and larger width. Each of these peaks consists of another two sub-peaks, corresponding to transitions from one SHF substate of the parent state to the same SHF substate of the daughter state. Two of the four SHF substates, respectively, are lying close enough to each other to have a frequency difference smaller than the laser linewidth ($\sim$100~MHz) and the Doppler width ($\sim$300~MHz) and can thus not be resolved while the other two lines have a splitting in the range of the broadening and thus result in a smaller and broader peak.

The measurements were all performed with a delay time $T$ between the two lasers of 350~ns and a target pressure of 250~mbar. These parameters shall provide a first comparison with results in $\overline{\text{p}}^4$He$^+$. A study at different laser delay times and target pressures is planned to be done in the future.
\linebreak

\begin{figure}[h!]
\centering
\includegraphics[scale=0.35, trim=275 100 325 175, clip]{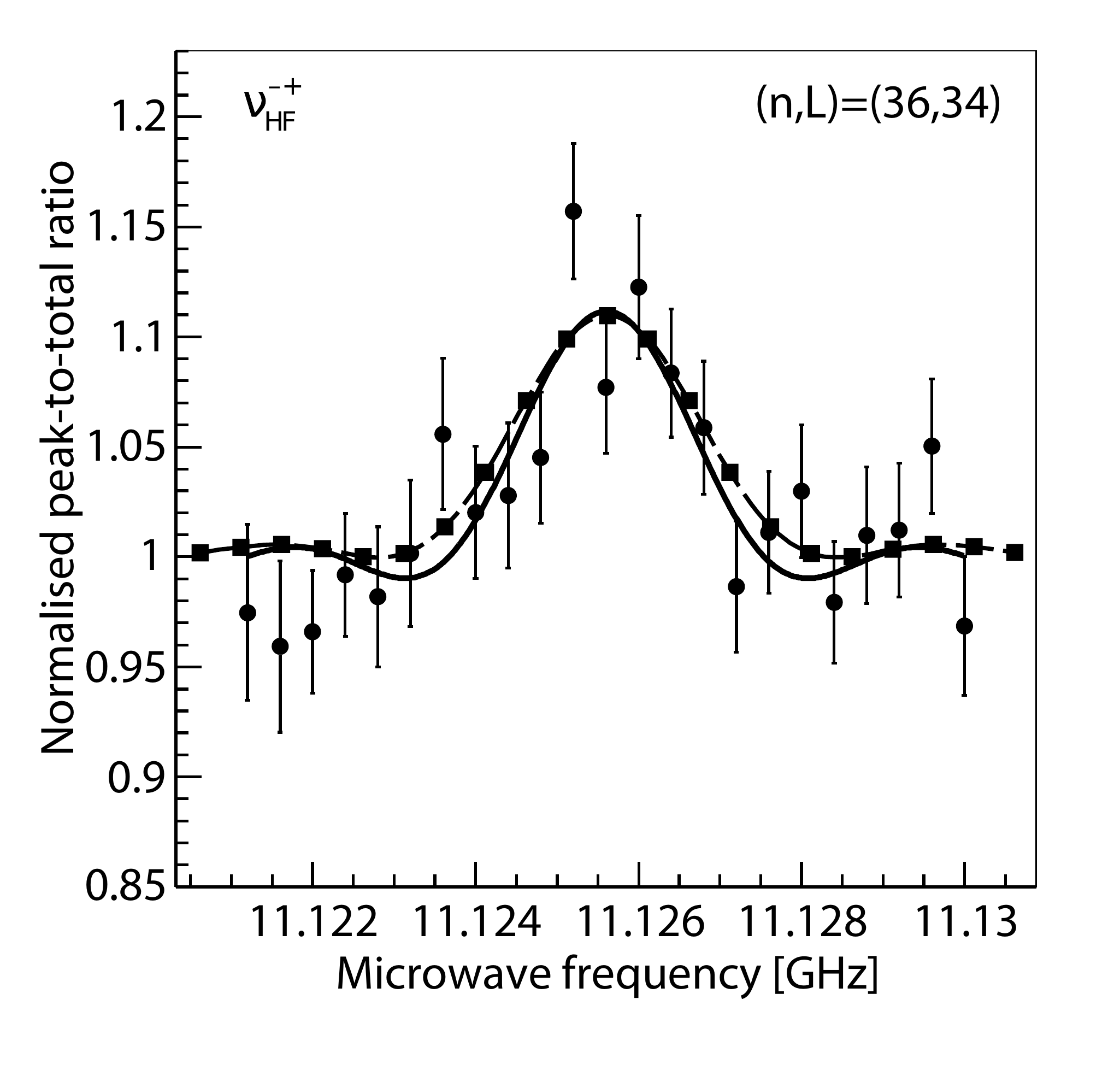} 
\includegraphics[scale=0.35, trim=275 100 325 175, clip]{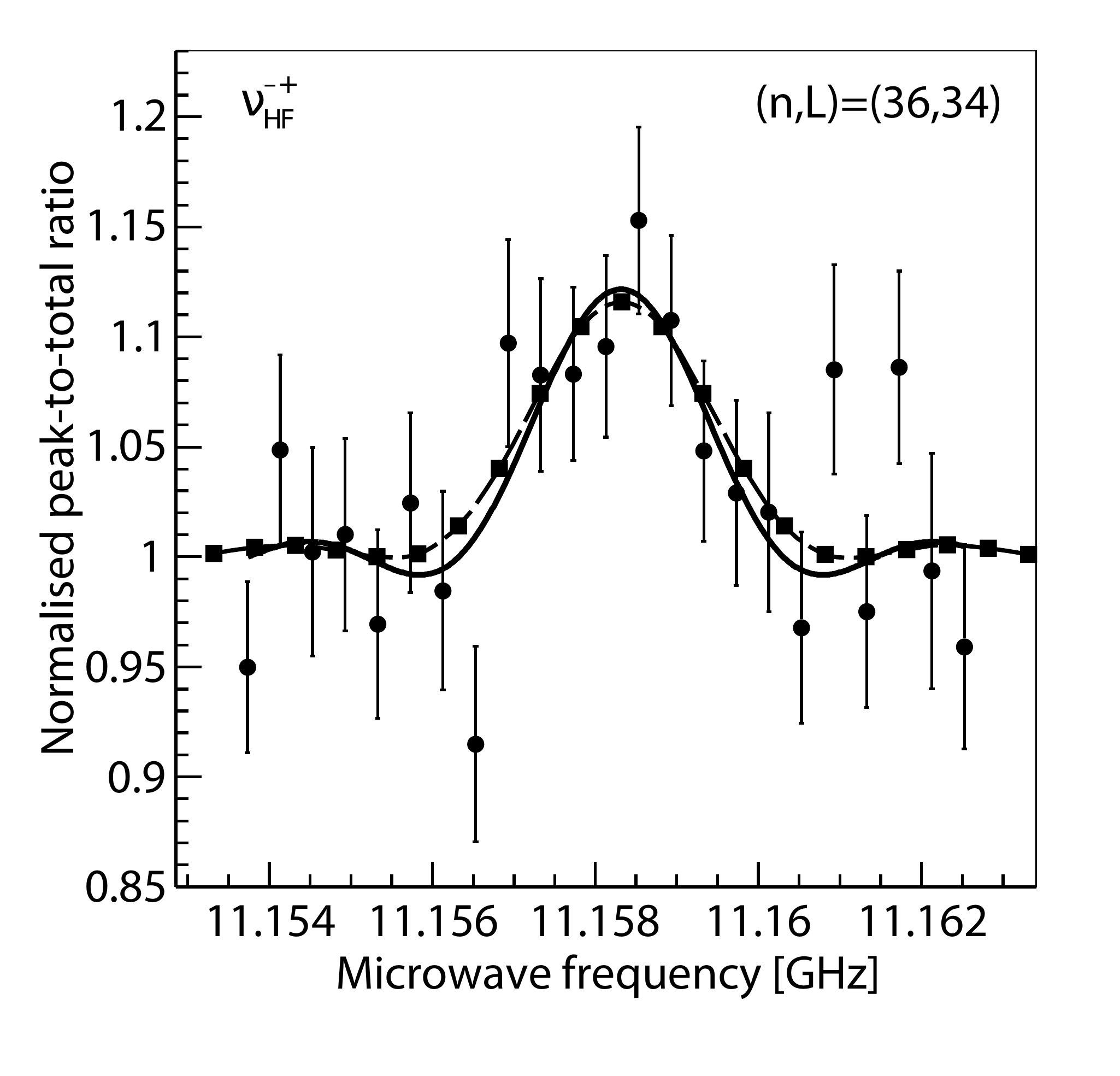} 
\caption{\small{Scan over the microwave frequency for two of the four SSHF transitions for the $(n,L)=(36,34)$ state of $\overline{\text{p}}^3$He$^+$, at a target pressure of 250~mbar. Each transition is fitted with Eq.~\ref{eq:NatLineShape} (solid lines). The frequencies of the measured transitions are 11.12559(14)~GHz and 11.15839(18)~GHz. The dashed curve shows a simulation using collision rates obtained from comparison between experiment and simulation.}}
\label{fig:MWscan}
\end{figure}

Two of the four "allowed" SSHF resonance transitions in $\overline{\text{p}}^3$He$^+$ could be observed. The scans of the two microwave-induced transitions are displayed in Fig.~\ref{fig:MWscan}. They were both fitted with the function of their natural line shape. For a two-level system, which is affected by an oscillating magnetic field for a time $T$, the line shape is given by~\cite{Sflugge}
\begin{eqnarray}    
\begin{split}                                                          
X(\omega) = A &\frac{\vert 2b \vert ^2} {\vert 2b \vert ^2 + (\omega_0 - \omega)^2} \\
\times &\sin^{2} \left\lbrace \frac{1}{2} \left [ \vert 2b \vert ^2 + (\omega_0 - \omega)^2 \right ] ^{\frac{1}{2}} T \right\rbrace.
\label{eq:NatLineShape}         
\end{split}
\end{eqnarray}
Here $X(\omega)$ is the probability that an atom is transferred from one HF state to the other, $\omega$ is the angular frequency of the magnetic field and $\omega_0$ is the angular frequency of the transition between the two energy levels. $A=1$ in an ideal two-level system. Thus $A$ is a scaling term added for the fitting procedure. It takes into account the fact that the real system is not an ideal two level one. The parameter $b=\Omega/2$ is a time independent part of the transition matrix elements between two energy levels, with the Rabi frequency $\Omega$. In the case of a complete $\pi$-pulse, one obtains $\vert b \vert T=\pi/2$. This is referred to as the optimum case, since together with $X(\omega)=1$ at resonance this gives the smallest width for the transition line, $\Gamma=\frac{0.799}{T}$~\cite{Sflugge}. The Fourier transform of the rectangular microwave pulse gives a lower limit for the transition line width. 
\linebreak

From the fit, the frequencies for the measured $\nu_{\text{HF}}^{--}$ and $\nu_{\text{HF}}^{-+}$ transitions can be obtained. All relevant results are summarized in Table~\ref{tab:Results}. For a pulse of length $T=350$~ns the expected width is $\Gamma=2.28$~MHz~\cite{Sflugge}, which is roughly in agreement with the measurement. In order to determine the optimum power to induce an electron spin flip and thus the maximum population transfer between two SSHF states, the signal was tested at several different microwave powers. The $\nu_{\text{HF}}^{--}$ transition was measured with a power of about 10~W and the $\nu_{\text{HF}}^{-+}$ transition at about 7.5~W. The magnetic field is depending on the microwave power through the relation 
\begin{eqnarray}    
B_0 = \sqrt{\frac{\mu_{0} Q P}{\pi a^2 d \omega J_{1}^{'2}(p_{11})}}
\label{eq:bfield}         
\end{eqnarray}
for the magnetic field in a cylindrical cavity with radius $a$ and length $d$. The resonance frequency is denoted $\omega$, $J_{1}^{'2}(p_{11})$ is a first root of Bessel function $J_{1}(x)$, $P$ is the microwave power and $Q$ is the measured value for the coupling factor of the cavity. This leads to an average oscillating magnetic field amplitude $B_0$ of 0.24(4)$\times 10^{-4}$~T and 0.19(3)$\times 10^{-4}$~T, respectively, inside the microwave cavity. The Rabi frequency $\Omega=\frac{\mu B_0}{\hbar}$, with $\mu$ denoting the calculated averaged magnetic dipole moment, is dependent on the microwave power. Using the values for the magnetic field and the magnetic dipole moment, we obtain a Rabi frequency in the range of 10~MHz for both powers. This is also in agreement with the results using the relations $\vert b \vert T=\pi/2$ and $b=\Omega/2$. The microwave power has to be chosen to achieve a complete $\pi$-pulse. Due to lack of time for more accurate power studies the two transitions have been measured at only one power each, which are slightly different from each other. These measurements will have to be repeated with improved statistics. 
\linebreak

There were also systematic effects, which had to be considered. The largest influence was due to the shot-to-shot fluctuations of the antiproton beam. These effects were reduced by normalising to the total intensity of the pulse and further normalising the second annihilation peak to the first one (refer to Section~\ref{sec:method}). Therefore mainly shot-to-shot fluctuations of the microwave power and deviations in the laser position and fluence from day to day -- although considerably smaller -- contributed to the error quoted in Table~\ref{tab:Results}. These contributions cannot be assessed individually. They are contained in the error obtained from the fit.
\begin{center} 
\begin{table*}
\caption{The first experimental results for the $\nu_{\text{HF}}^{--}$ and $\nu_{\text{HF}}^{-+}$ in comparison with three-body QED calculations, where $\nu_{\rm{\text{HF}}}$ denote the SSHF transition frequencies, $\delta_{\text{exp}}$ is the relative error of the measured frequencies and $\Gamma$ the resonance line width. The theoretical precision is $\sim$5$\times 10^{-5}$.}\label{tab:Results}
{\small
\hfill{}
\begin{tabular}[t!]{|r@{}|r@{}|r@{}|r@{}|r@{}|r@{}|r@{}|r@{}|r@{}}
\hline
\hline
\textbf{}& \textbf{$\nu_{\text{exp}}$ (GHz)}& \textbf{$\delta_{\text{exp}}$ (ppm)}& \textbf{$\Gamma$ (MHz) }& \textbf{Korobov~\cite{Korobov:06,Korobov:2010}}& \textbf{$\Delta \nu_{\text{th-exp}}$ (ppm)}& \textbf{Kino~\cite{Kino:03APAC}}& \textbf{$\Delta \nu_{\text{th-exp}}$ (ppm)} \\
\hline
\textbf{$\nu_{\text{HF}}^{--}$ (GHz)} & 11.125 59(14) & 13 & 2.08(22) & 11.125 00(55) & -53 & 11.125 15(55) & -39 \\
\textbf{$\nu_{\text{HF}}^{-+}$ (GHz)} & 11.158 39(18) & 16 & 1.92(19) & 11.157 73(55) & -59 & 11.157 56(55) & -74 \\
\hline
\textbf{$\Delta \nu_{\text{HF}}^{\pm}$ (GHz)} & 0.03279(22) &  &  & 0.0327219(16) &  &  &  \\
\hline
\hline
\end{tabular}}
\hfill{}
\end{table*}
\end{center}
The experiment has been numerically simulated by solving the optical Bloch equations in order to estimate important measurement parameters, in particular the required microwave power and the signal-to-noise ratio. The Bloch equations describe the depopulation of states, in this experiment induced by laser light and microwave radiation and under the influence of collisional effects. For most parameters, such as microwave power, $Q$ value and laser delay, the measured values were taken. The rates of collisional effects -- inducing relaxations between the SSHF states -- were obtained from adjusting the simulation to the experimental results for the $\nu_{\text{HF}}^{--}$ resonance, and they are comparable to the values for $\overline{\text{p}}^4$He$^+$. Fig.~\ref{fig:MWscan} shows the results of the simulations in comparison to the fitted measurement data. The numerical simulations are explained in detail in~\cite{Pask:2008p}.  
\linebreak

The measured hyperfine transition frequencies agree with theory~\cite{Korobov:06} within less than 1~MHz. The current precision of $\sim$20~ppm is still worse than for the most recent results with $\overline{\text{p}}^4$He$^+$, which gave an error of 3~ppm for the individual transition lines~\cite{Pask:09}. Due to limitations in antiproton beam quality this precision for $\overline{\text{p}}^4$He$^+$ is not likely to be improved anymore. However, it is also unlikely to achieve an error for $\overline{\text{p}}^3$He$^+$ as small as for $\overline{\text{p}}^4$He$^+$. There are eight instead of four SSHF energy levels in $\overline{\text{p}}^3$He$^+$ and thus the measured signal will be only about half of the signal obtained for $\overline{\text{p}}^4$He$^+$. Therefore much higher statistics would be required.
\linebreak

A comparison of the theoretical values for the two SSHF transitions at 11~GHz with the measurement results shows that there is a small shift in frequency towards higher values for both transitions. The frequency difference $\Delta \nu_{\text{th-exp}}$ between theory and experiment is $\sim$0.6~MHz for $\nu_{\text{HF}}^{--}$ and $\sim$0.7~MHz for $\nu_{\text{HF}}^{-+}$ respectively. According to V. Korobov~\cite{Korobov:2010}, this discrepancy is most likely due to the theoretical limits of the Breit-Pauli approximation that has been used for the calculations. The relative error of the theoretical frequencies is estimated to be $\alpha^2=5 \times 10^{-5}$. The theoretical error for the frequency difference between theory and experiment would then be $\sim$0.6~MHz. Together with the experimental error of $\sim$0.2~MHz there is agreement between experiment and theory.
Higher order correction terms need to be calculated to improve the theoretical results. The work on these calculations is in progress~\cite{Korobov:2010}. 
\linebreak

A density dependent shift could also contribute to this deviation. The density dependence is found to be much smaller for the M1 transitions, the electron spin flip transitions induced by the microwave, than for the E1 transitions induced through laser stimulation~\cite{Pask:08}. For $\overline{\text{p}}^4$He$^+$ theoretical calculations of Grigory Korenman~\cite{Korenman:2006,Korenman:2009} confirmed that the density dependence is very small. Also for $\overline{\text{p}}^3$He$^+$ theory predicts a collisional shift much smaller than our error bars~\cite{Korenman:2010}. 
\linebreak

The deviation between the experimental and theoretical values for the frequency difference $\Delta \nu_{\text{HF}}^{\pm} = \nu_{\text{HF}}^{-+} - \nu_{\text{HF}}^{--}$ between the two SSHF lines at 11~GHz is 68~kHz out of 32 MHz. This difference is important due to its proportionality to the magnetic moment of the antiproton. The error of the theoretical value for $\Delta \nu_{\text{HF}}^{\pm}$ is 1.6~kHz, which is considerably smaller than the error of 220~kHz for the value obtained from the measured transitions. The reason is that in theory the splitting between the transition lines can be calculated directly and the errors are the same for all transitions within the hyperfine structure whereas the experimental value of the splitting is received from the difference of the single transition lines. Even though the experimental error is much larger than the theoretical one, there is agreement between theory and experiment within their errors.
\end{linenumbers}

\section{Conclusions}
\begin{linenumbers}
We have reported on the first microwave spectroscopic measurement of the hyperfine structure of $\overline{\text{p}}^3$He$^+$. 

Two of the four favoured SSHF resonance transitions in $\overline{\text{p}}^3$He$^+$ were observed and are in agreement with theory~\cite{Korobov:06} within the estimated theoretical error. Also the frequency difference $\Delta \nu_{\text{HF}}^{\pm}$ agrees with theoretical calculations. However, the experimental error for $\Delta \nu_{\text{HF}}^{\pm}$ is still very large compared to theory.

A systematic study of these transitions and improved statistics will allow a higher precision, in particular for the frequency difference between the SSHF transition frequencies, for which the experimental error is still considerably larger than the theoretical error. It is further planned to measure also the two SSHF transitions at 16~GHz. The density dependence should be tested and different laser delay timings and microwave powers will be studied. 
\end{linenumbers}

\section{Acknowledgements}
{We are grateful to Dr. V. Korobov and Dr. G. Korenman for fruitful discussions on the theoretical framework. Further, we want to thank especially Ms. Silke Federmann and project student Mr. Mario Krenn for their help before and during the beamtime. This work has received funding from the Austrian Science Fund (FWF): [I--198--N20] as a joint FWF--RFBR (Russian Foundation for Basic Research) project, the Austrian Federal Ministry of Science and Research, the Japan Society for the Promotion of Science (JSPS), the Hungarian National Science Funds (OTKA K72172), the European Science Foundation (EURYI) and the Munich Advanced Photonics Cluster (MAP) of the Deutsche Forschungsgemeinschaft (DFG).}

\end{document}